# Deterministic Bottom-Up Fabrication of Plasmonic Nanostructures on Optical Nanofibers via Blurred Electron Beam Deposition

**Antonio Balena[1,*], Marianna D'Amato[1], Muhammad Fayyaz Kashif[2,3], Chengjie Ding[1], Massimo De Vittorio[3,4], Ferruccio Pisanello[3], Alberto Bramati[1]**

[1] *Laboratoire Kastler Brossel, Sorbonne Université, CNRS, ENS-PSL Research University, Collège de France, France*
[2] *Università degli Studi di Napoli Federico II – Department of Electrical Engineering and Information Technologies, Italy*
[3] *Istituto Italiano di Tecnologia - Center for Biomolecular Nanotechnologies, Italy*
[4] *Technical University of Denmark - Department of Health Technology Drug Delivery and Sensing, IDUN Section, Denmark*

[*] *Corresponding author:* antonio.balena@lkb.upmc.fr

E-mail addresses:
Marianna D'Amato: marianna.damato@lkb.upmc.fr
Muhammad Fayyaz Kashif: muhammadfayyaz.kashif@unina.it
Chengjie Ding: dcjnonno@gmail.com
Massimo De Vittorio: massimo.devittorio@iit.it
Ferruccio Pisanello: ferruccio.pisanello@iit.it
Alberto Bramati: alberto.bramati@lkb.upmc.fr

## 1   Abstract

This study introduces a novel method for the deterministic fabrication of metallic nanostructures with controlled geometry and composition on suspended, single-mode tapered optical nanofibers (TNFs) using a tailored Blurred Electron Beam Induced Deposition (BEBID) technique. TNFs, owing to their subwavelength diameters and intense evanescent fields, offer a unique platform for enhanced light–matter interactions at the nanoscale. However, their mechanical fragility has thus far hindered the integration of plasmonic structures using conventional high-energy deposition methods. BEBID addresses this limitation by deliberately defocusing the electron beam to reduce local mechanical stress, minimize vibration, and prevent fiber damage during deposition, thereby enabling the one-step growth of platinum nanopillars with sub-20 nm spatial precision and high structural fidelity directly on suspended TNFs. The fabricated structures were characterized using SEM, EDX, and their optical properties were investigated through broadband scattering spectra and polarization-resolved measurements, showing strong agreement with Finite-Difference Time-Domain (FDTD) simulations. Numerical modeling further reveals that ordered arrays of nanopillars can shape and direct the scattered field along the fiber axis, enabling directional emission. This work establishes BEBID as a versatile bottom-up nanofabrication approach for functional photonic architectures on fragile substrates, with direct applications in quantum photonics, nano-optics, and on-fiber plasmonic sensing.

## 2   Introduction

For decades, the emergence of localized surface plasmon resonances (LSPRs) resulting from the interaction of electromagnetic fields with subwavelength-sized metal structures has been extensively studied[1]. These phenomena have been exploited in a wide range of applications, including sensing and biosensing[2,3], neuroscience for controlling neural activity[4–6], optical tweezers and micro/nano manipulation [7,8], and quantum photonics, thanks to the demonstration of single-photon emission enhancement by plasmonic nanocavities[9–11]. Recently, integrating plasmonic nanostructures with multimode fiber optics has gained significant attention, opening new opportunities such as spatially-resolved[12] and wide-volume[13,14] surface-enhanced Raman scattering (SERS), light shaping of emission and collection[15], biosensing[16], and actuation[17]. While these advances highlight the promise of combining plasmonic structures with optical fibers for scientific and technological innovation, they are mostly focused on systems where thousands of guided or radiative modes interact with thousands of plasmonic nanostructures. Such configurations generate broadband responses and prevent studying or controlling light-matter interactions at the single-nanostructure level.

To push applications toward the quantum regime, it would be essential to achieve controlled interaction between a single mode and a single structure, an interaction regime that remains largely unexplored due to the lack of fabrication technologies able to nucleate single plasmonic nanostructures within single-mode photonic systems deterministically. In this context, high-transmission single-mode tapered optical nanofibers (TNFs)[18] – optical fibers mechanically tapered to sub-wavelength diameters – represent a promising photonic platform. They support a highly controlled and deterministically defined intense evanescent field near the fiber surface, ideal for strong interactions with the external environment or nanoparticles deposited on them[19–21], as well as for quick and stable connection with fiber-based platforms. Integrating LSPR onto TNFs, therefore, holds great potential both for (i) applications, where the combination of plasmon-enhanced scattering and absorption with the strong evanescent field can enhance LSPR-based sensing or detection of refractive index changes, and (ii) fundamental studies in nonlinear optics[22,23], cavity quantum electrodynamics[24,25], and single photon emission enhancement[26] that may benefit from the extreme confinement of the electromagnetic energy.

Despite the potential, currently available technologies only allow these investigations to be conducted on non-deterministic coupling conditions, and deterministic plasmonic integration on TNFs has remained a significant challenge. Sugawara et al.[27] and Shafi et al.[28] recently achieved integration of gold nanorods and single photon emitters on TNFs with stochastic coupling conditions, showing how LSPR-mediated single-photon emission enhancement is significantly influenced by the nanostructures' geometry and orientation on the emission diagram of the quantum emitters. From a fabrication standpoint, most approaches rely on self-assembly, which lacks precision in nanoparticle's placement, spacing, and orientation, all critical aspects for exploiting LSPRs effectively. In addition, being inherently dielectric and nanometric in diameter, TNFs are often suspended over a few centimeters, making them highly sensitive to mechanical stress and damage during fabrication. This has been a major roadblock to the deterministic integration of functional plasmonic nanostructures on TNFs. Conventional techniques such as high-energy electron beam lithography and chemical vapor deposition exert excessive mechanical or thermal stress, leading to fiber deformation or breakage. An alternative approach uses thin films that cover a large part of the TNF nanometric section to exploit surface plasmon resonances (SPR), but the interaction region is extended along all the nanofiber, thus lacking the spatial localization provided by discrete plasmonic elements such nanopillar or nanoantennae.

All these observations underscore the need for methods that can deterministically nucleate LSPR directly onto the TNFs without mechanical, optical, or thermal alterations, and with nanometer precision. Addressing this fabrication bottleneck is the primary motivation for this work. To this aim, we present a novel method for fabricating controlled-geometry platinum (Pt) nanopillars and nanoantennae deterministically placed onto TNFs with sub-20 nm resolution, achieved using an unconventional adaptation of Electron Beam Induced Deposition (EBID)[29,30]. The TNF's narrowest section was suspended, kept under axial traction, and exposed under the SEM beam, while an organometallic platinum precursor was introduced into the chamber. In contrast to the established Focused-EBID (FEBID), our approach deliberately exploits beam defocusing to reduce the local pressure exerted by the electron beam on the suspended, nanometric, and dielectric TNF, which could be otherwise put in vibration by a tightly focused electron beam, hindering reliable fabrication. We refer to this method as *blurred* EBID (BEBID). This technique enabled the nucleation of one or multiple nanostructures on a TNF, which were optically characterized by recording the broadband spectral dispersion of scattered light and by polarization measurements. The results are in satisfactory agreement with Finite-Difference Time-Domain (FDTD) numerical simulations and confirm the efficacy of the BEBID technique. Numerical simulations also demonstrate the impact of ordered arrays of Pt nanopillars on TNFs, predicting increased overlap between the scattered electromagnetic field and the guided modes of the fiber. We anticipate that this approach can be extended to enable the fabrication of increasingly complex nanostructures for plasmonic light manipulation in TNFs, opening new avenues for TNFs' application across various scientific fields, including sensing and quantum photonics, but also to build functional plasmonic platforms on fragile photonic systems.

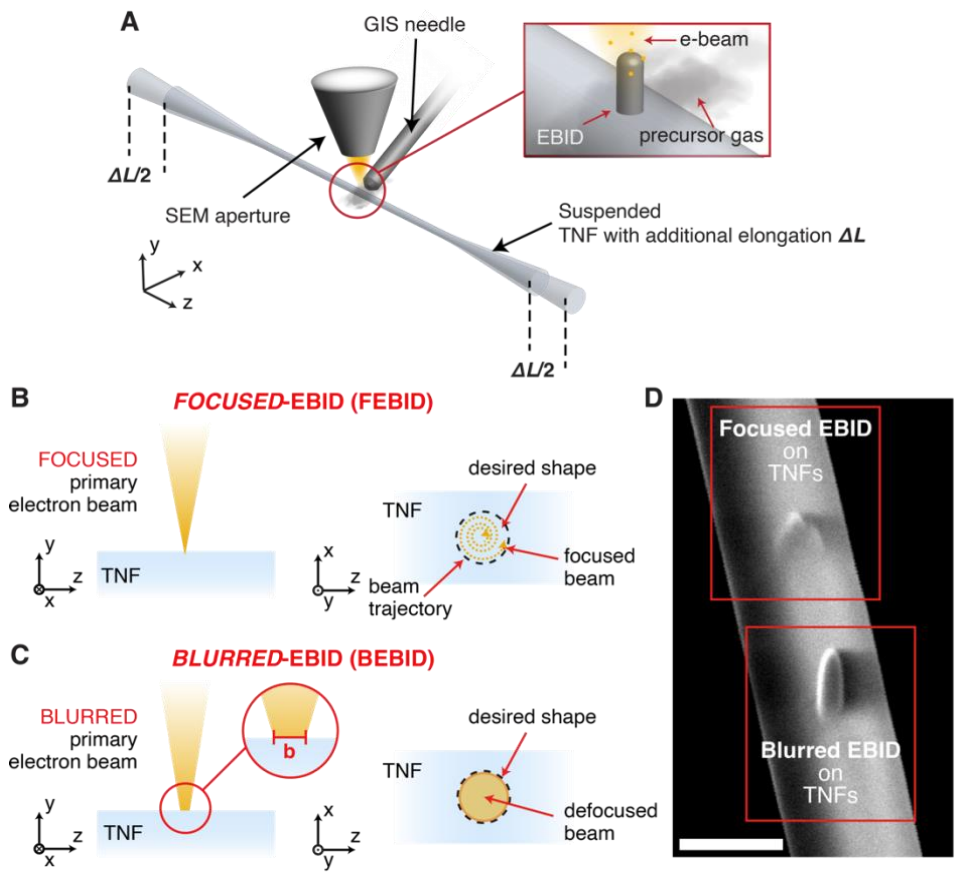

*Figure 1. Blurred Electron Beam-Induced Deposition. A) Sketch of the EBID fabrication of nanostructure on TNFs. During the fabrication, the fiber is suspended, and the nanometric waist region is focused under the electron beam. B) Schematized description of the FEBID process. The beam is tightly focused, and the focal spot is moved along a spiral trajectory over the sample surface.*

*C) Schematized description of the BEBID process. The focal spot is moved along the y-axis so that the resulting blurring "b" is usually set at the diameter of the desired nanopillar. D) SEM micrograph of two pillars fabricated on the same TNF following the same process except for the beam blurring. The top pillar is realized through FEBID and results in an undefined shape, the bottom one is deposited through BEBID, and a pillar shape with a base diameter of 100 nm and a height of 300 nm is obtained. The scale bar is 300 nm. To enhance visibility, the nanofiber is rotated on the xy-plane and tilted by 52° with respect to the z-axis.*

## 3 Results

In a subwavelength diameter TNF, the boundary conditions of guided light are influenced by the large refractive index contrast between the nanofiber and the surrounding air[31], in contrast to conventional optical fibers. Consequently, a significant evanescent field extending beyond the nanofiber's boundaries is observed, extinguishing into the surrounding air over a few hundred nanometers, with a substantial portion of the light propagating at the fiber's surface. The electromagnetic field around the TNF is determined by factors such as the propagation mode, the wavelength of the guided light, and the geometry and refractive index of the TNF itself and the surrounding environment[32]. The **Supplementary Material** and **Supplementary Figures S1-S4** report the detailed design of the TNFs used in this work, while the details on the fabrication of the TNFs are reported in the "Materials and Methods" section, and **Supplementary Figures S5-S9**.

Here we strive at nucleating single LSPR on single-mode TNFs where only the fundamental mode $HE_{11}$ propagates, a particularly challenging task due to the sub 400 nm diameter of the TNFs, their purely dielectric nature, and their suspended geometry. Many conventional nanofabrication strategies are unsuitable to achieve this aim in a deterministic fashion. For instance, the diameter of TNFs is below the diffraction limit of imaging systems typically used in photolithography, while immersion into viscous photoresist or electron beam lithography (EBL) resist would likely cause the fiber to break. Although top-down approaches, such as Focused Ion Beam Milling[33] (FIBM) using both gallium ions[34] and helium ions[35] or also laser ablation by femtosecond pulses[36], have been successfully used to pattern TNFs, to the best of our knowledge, no bottom-up approaches for deterministic fabrication of metallic nanostructures on TNFs have been reported so far.

To overcome these limitations, we have chosen Electron Beam Induced Deposition (EBID). EBID allows the precise deposition of nanostructures by using an electron beam to locally dissociate a gaseous precursor material on the surface of the substrate. TNFs were loaded into a dual configuration Focused Ion Beam/Scanning Electron Microscopy (FIB/SEM) system (FEI Helios NanoLab 600i) equipped with a three-channel gas injection system (GIS). The TNF was suspended with a controlled axial traction applied, while exposed to the electron beam (**Figure 1.A**). The platinum precursor used was MeCpPtMe$_3$ locally injected through the nozzle at the upper side of the sample. A specific aspect of the fabrication process was the deliberate blurring of the electron beam, achieved by controlled defocusing during exposure, to minimize the mechanical stress exerted on the fiber under the high electron flux of standard focused EBID (FEBID, **Figure 1.B**). We refer to this technique as *blurred EBID* (BEBID, **Figure 1.C**). While typically considered detrimental to nanofabrication, this beam blurring enabled successful patterning on these specific substrates while maintaining the composition of the nanostructure unchanged[37]. The beam blurring is controlled by the parameter *b*, which determines the diameter of the beam at the sample plane and was adjusted to match the base diameter of the plasmonic structure to be nucleated (a nanopillar), resulting in a defocusing of approximately 3 μm below the sample surface. The tilted SEM micrograph in **Figure 1.D** shows a comparison of a successful BEBID single pillar nucleation (100 nm base diameter, 300 nm height) on a TNF with a 187.5 nm radius of curvature, compared with a FEBID process with the same nominal parameters.

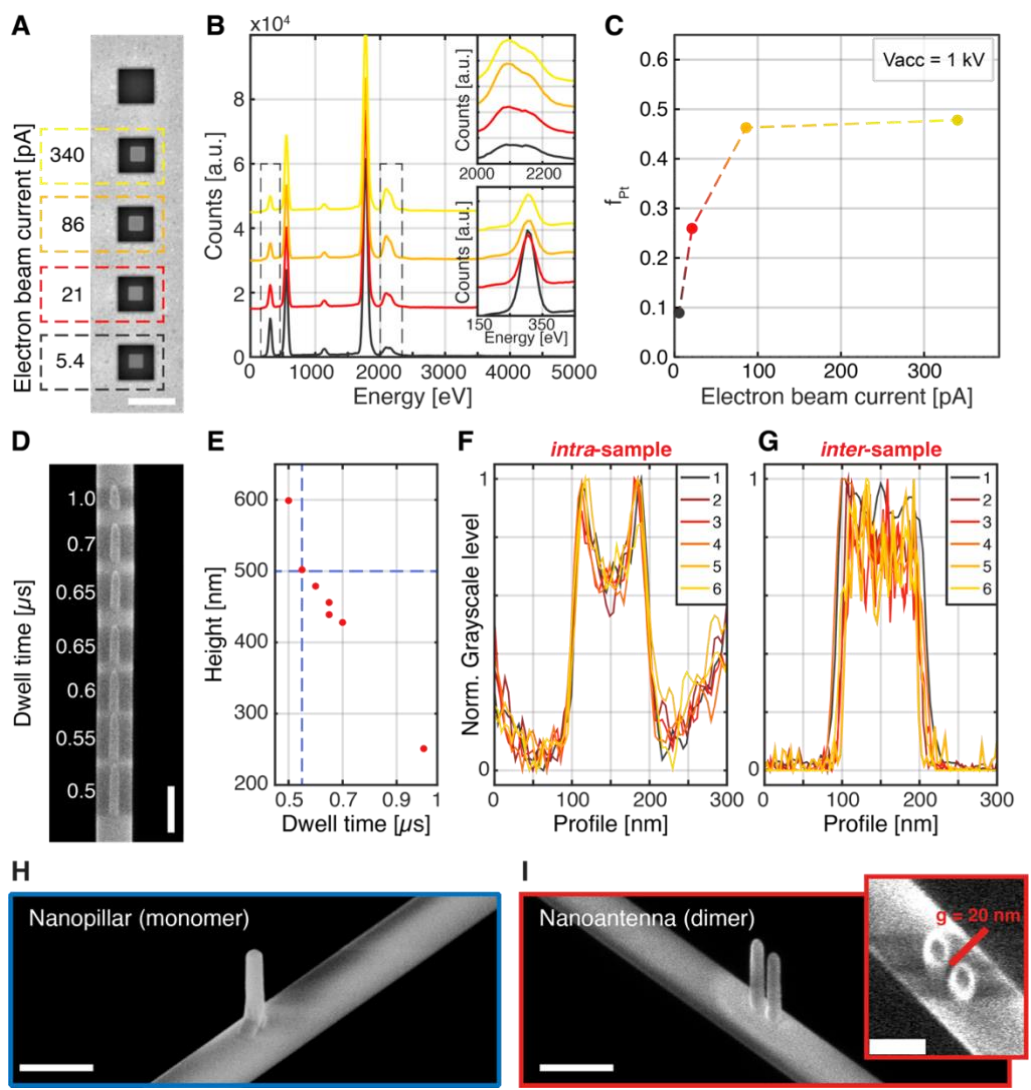

*Figure 2. Fabrication of BEBID Nanostructures. A) SEM micrograph of the sample fabricated for EDX characterization. A series of 800x800x500 $nm^3$ cubes at the center of $SiO_2$ pads were obtained from Focused Ion Beam Milling through a 50 nm-thick gold layer and consecutive EBID. The scale bar is 3 μm. B) EDX spectra measured from the cubes in Panel C. The insets show the spectral regions corresponding to (top) the Pt Mα x-ray line and (bottom) the C Kα x-ray line. C) Fill factor $f_{Pt}$ of Pt to the total volume of Pt+C extracted from the spectra in Panel D. D) Tilted SEM micrograph of a series of pillars fabricated on a TNF by setting different dwell time values. The scale bar is 500 nm. E) Height of the pillars measured by SEM inspection VS dwell time. The horizontal dashed line indicates the target height h = 500 nm, while the vertical one indicates the corresponding dwell time of 0.55 μs. ) Grayscale profiles of different pillars fabricated onto the same TNF to evaluate the intra-sample repeatability of the method to obtain a target diameter of 100 nm. G) Grayscale profiles of different pillars fabricated onto the same TNF to evaluate the inter-sample repeatability of the method to obtain a target diameter of 100 nm. H) SEM micrograph of a nanopillar with a base diameter of 100 nm and a height of 500 nm fabricated on a d = 360 nm TNF via BEBID. The scale bar is 500 nm. I) SEM micrograph of a nanoantenna constituted by two identical nanopillars with the same geometry shown in panel C. The scale bar is 500 nm. The inset shows the same nanostructure seen from the top, showing the nanogap separating the two pillars. The scale bar is 250 nm.*

Since EBID-deposited structures often contain significant carbon contamination due to incomplete dissociation of the precursor molecules, we investigated the composition of BEBID-deposited materials to optimize their plasmonic responses. **Figure 2.A** reports test depositions at different beam currents $I_{beam}$, at an accelerating voltage of $V_{acc}$ = 1 kV, since exposing the TNF at higher $V_{acc}$ could compromise its integrity. Energy-dispersive X-ray spectroscopy (EDX, spectra in **Figure 2.B**) allowed us to characterize the relative fill factor ($f_{Pt}$) of platinum to carbon as a function of $I_{beam}$. The results are plotted in **Figure 2.C**, suggesting that the composition of the structures can be tuned by adjusting the fabrication parameters and allowing the selection of an optimal balance between satisfying the low acceleration voltage requirements imposed by the substrate and maximizing the amount of Pt in the nanostructure. By setting $I_{beam}$ = 86 pA and $V_{acc}$ = 1 kV, we achieved a $f_{Pt}$ = 0.46. Subsequently, an empirical approach was used to determine the optimal dwell time for fabrication on TNFs. Using a target geometry of a 500 nm high pillar with a base diameter of 100 nm, a series of pillars were fabricated with dwell times $\tau_D$ varying from 0.5 µs to 1 µs (**Figure 2.D**). The dwell time that gives the best match between the target geometry and the actual fabricated geometry was experimentally determined at $\tau_D$ = 0.55 µs by SEM inspection, as reported in **Figure 2.E**.

To evaluate the repeatability of the process, SEM inspection was performed on multiple fabricated structures. In particular, **Figure 2.F** displays the profiles extracted from a series of six pillars fabricated onto the same TNF, used to assess *intra*-sample variability. This process yielded an average pillar diameter of $d_{avg}$ = 98.2 ± 0.9 nm (mean value ± standard deviation), corresponding to a relative error of approximately 1%, thereby demonstrating remarkable fabrication repeatability. **Figure 2.G** shows the profiles extracted from six pillars fabricated on six different TNFs. The profiles are used in this case to assess the *inter*-sample variability. In this case, the extracted average diameter of $d_{avg}$ = 104 ± 9 nm, with a relative error of around 8%. The higher standard deviation observed in the inter-sample analysis is attributed not to limitations of the fabrication process itself, but rather to uncontrolled variability introduced during manual preparation of the TNF. Specifically, the manual gluing of the nanofibers onto the sample holder may lead to non-uniform mechanical tension across samples, which in turn affects the final stability of the fiber during electron beam exposure. This suggests that the observed differences are extrinsic to the technique and could be minimized through improved fiber mounting procedures. Finally, the SEM micrograph in **Figure 2.H** demonstrates an example of how the technique can achieve the successful fabrication of single bottom-up nanopillars with a high aspect ratio on a TNF with a curvature radius smaller than the nanostructure height. Using the same technique, multiple structures can be fabricated sequentially, with well-controlled nanogaps separating them down to a few tens of nanometers, as the nanoantenna shown in **Figure 2I**.

Post-fabrication, the nanostructures-decorated TNFs underwent a plasma-oxygen exposure to cleanse the surface by removing potential dust particles (see **Supplementary Figure S10**), which could impair optimal coupling with the fabricated nanostructures[38]. Additionally, the plasma-oxygen treatment exerted a "purifying" effect on the composition of EBID metallic nanostructures, reducing the carbon content while enhancing the metal content, as reported in previous studies[39,40]. This effect was further characterized on test structures, where a diameter reduction of approximately 10 nm and a height reduction of about 25 nm was observed by SEM inspection (see **Supplementary Figure S11**). Since the volume reduction is indeed related to the decrease in the dielectric component of the nanostructures, an increase in the metal percentage from 47.5% to about 65% was calculated, given the fabrication parameters identified above and an initial nanostructure height of $h = 525$ nm and diameter of $d = 110$ nm.

The optical interaction between the evanescent field guided by the TNF and a target structure comprising a nanopillar with a base diameter of $d = 100$ nm and a circular cross-section (hereafter referred to as a "monomer") was characterized and compared to a nanoantenna configuration, consisting of two nanopillars separated by a $g = 20$ nm nanogap (hereafter referred to as a "dimer"). This observation is the basis for numerical modeling of the optical properties of the nanostructures since it warrants the calculation of the effective refractive index of the Pt-C composite by Bruggemann's effective medium approximation, for which the relative permittivity $\varepsilon_{eff}$ of the composite results[41]:

$$\varepsilon_{eff} = \frac{1}{2}\{-[\varepsilon_{Pt}(f_C - 2f_{Pt}) + \varepsilon_C(f_{Pt} - 2f_C)] \pm \sqrt{[\varepsilon_{Pt}(f_C - 2f_{Pt}) + \varepsilon_C(f_{Pt} - 2f_C)] + 8\varepsilon_C\varepsilon_{Pt}}\}, \quad (1)$$

where $\varepsilon_{Pt}$ and $\varepsilon_C$ are the permittivities of Pt and C, respectively, and $f_{Pt}$ and $f_C = 1 - f_{Pt}$ are the volume fractions of the two materials, and where the condition that drives the choice of the sign for the square root term is Im$\{\varepsilon_{eff}\} \geq 0$. **Supplementary Figure S12** reports the calculated $\varepsilon_{eff}(\lambda = 785$ nm) for $f_{Pt} \in [0.6, 1]$.

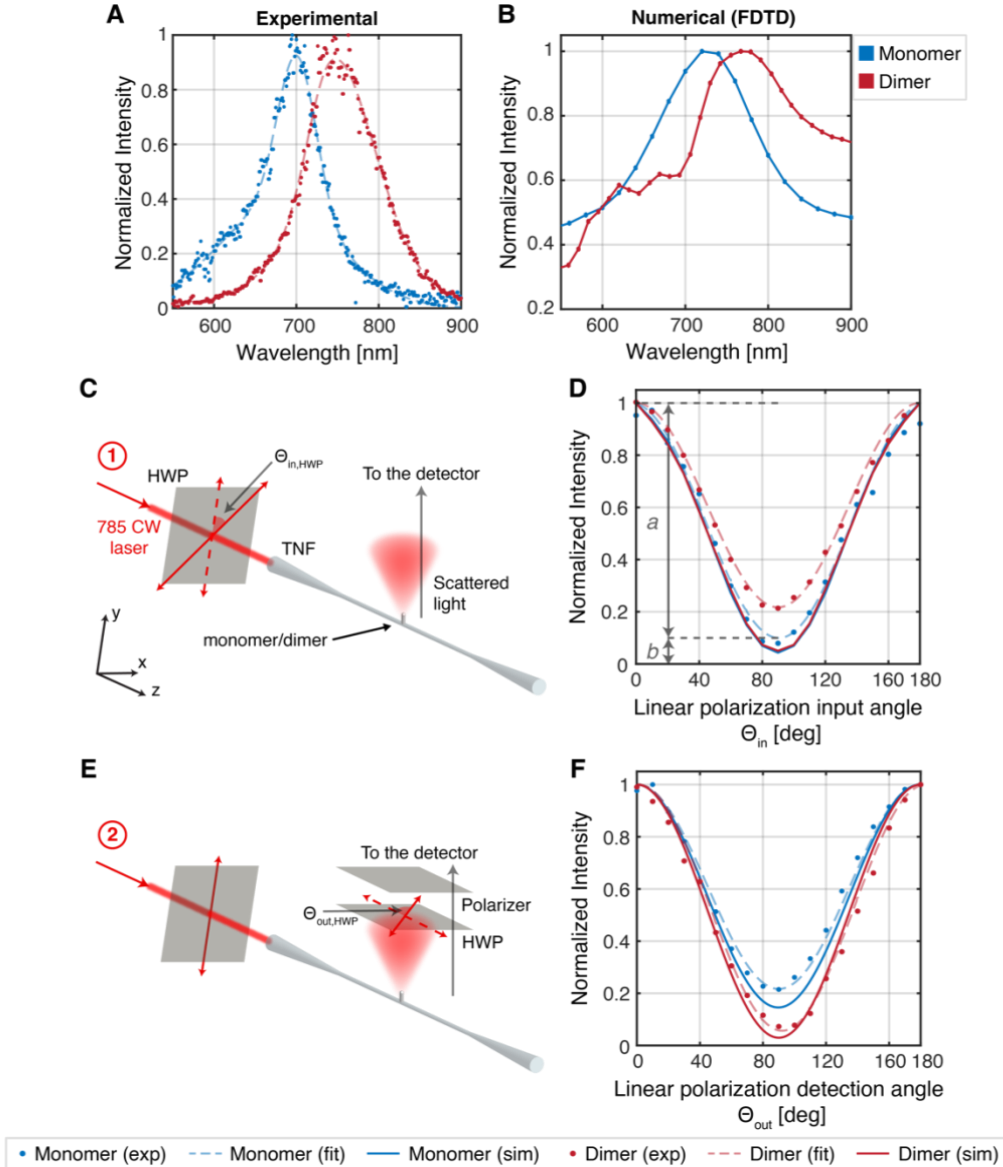

*Figure 3. Optical characterization. A) Normalized experimental broadband extinction spectra of the monomer (blue) and the dimer (red). Dashed lines are meant as a guide for the eyes. B) Normalized numerical extinction spectra of the monomer (blue) and the dimer (red). C) Sketch of the measurement configuration used for the dependence of the scattering intensity from the input linear polarization characterization. D) Normalized polar plots of the scattering from a monomer (blue data points and curves) and a dimer (red data points and curves) measured for the configuration in panel C. Circles are the experimental data, the dashed line is the fit over the experimental data, and the continuous line is the numerical curve. The arrows labeled 'a' and 'b' refer to the Malus' law-like formula used for data fitting. E) Sketch of the measurement configuration used for estimating the DoLP of the scattered light. F) Normalized polar plots of the scattering from a monomer (blue data points and curves) and a dimer (red data points and curves) measured for configuration in panel E. Circles are the experimental data, the dashed line is the fitting over the experimental data, and the continuous line is the numerical curve.*

The spectral position of the resonance peak for scattering was measured experimentally on both monomer and dimer nanostructures. A supercontinuum laser was coupled to one end of the TNF, and the scattered light from the nanostructures was collected by an objective lens and directed to a spectrometer. As shown in the spectra in **Figure 3.A**, the dimer exhibits a redshift of 50 nm relative to the monomer, with the peak shifting from $\lambda = 700$ nm to $\lambda = 750$ nm. To validate the experimental data, the scattering spectra of the monomer and dimer were also simulated numerically using a Finite Difference Time Domain model (FDTD), as illustrated in **Figure 3.B.** The monomer spectrum reveals a peak in the near-infrared at approximately $\lambda = 720$ nm, while the dimer spectrum displays a redshift of 47 nm, with a peak at $\lambda = 767$ nm, aligning well with the experimental data. The redshift can be attributed to the shaping effect on LSPR in the case of two closely packed, identical nanostructures. A small interpillar distance induces a significant enhancement of the electromagnetic field in the gap region between the nanostructures. Additionally, a reduced interpillar distance fosters intense mutual interaction, leading to coupling between modes of different orders. As the interpillar distance decreases, higher-order resonances emerge. The primary consequence of this interaction is the redshift of resonance in the dimer case, which manifests as a hybridized mode resulting from the convolution of multiple modes into a single and broad band. Indeed, both the numerical and experimental spectra show spectral broadening between the monomer and dimer, in addition to the redshift. In the experimental case, this broadening is observed as an increase from $FWHM_{monomer} = 76$ nm to $FWHM_{dimer} = 97$ nm. Numerically, the broadening is also evident, though exact estimation is less straightforward. The spectra obtained from the simulations are broader than those of the experiments because, in the simulations, a spectrally flat source over the explored wavelength range is assumed to be injected directly into the nanometric section of the TNF, close to the nanostructure. In the experiment, the spectrum of the supercontinuum laser injected into the fiber, endowed from the beginning with a given spectral dispersion, passes through the tapered section of the TNF, whose profile, optimized for adiabatic transmission in a range around a wavelength of 785 nm, acts as a filter for wavelengths outside that range.

Thereafter, two experimental configurations were employed to investigate the polarization dependence of the scattering properties of the two distinct types of TNF-coupled BEBID nanostructures. The first configuration (**Figure 3.C**) was used to measure the dependence of scattered light intensity on the input polarization, while the second configuration (**Figure 3.E**) was utilized to determine the polarization state of the light scattered by the nanostructures when the input beam polarization was fixed. Both configurations have been numerically simulated using an FDTD model. As sketched in **Supplementary Figure S13**, a laser with a wavelength of 785 nm (Toptica Photonics), close to the resonance peak of the dimer, was directed through a half-wave plate (HWP) before being coupled into one end of the TNF. This latter was maintained in an unbent state to prevent the introduction of asymmetric stress on its cross-section, thereby preserving the polarization of the propagating light by minimizing birefringence.

Scattering from the nanostructure was monitored using an sCMOS (Scientific Complementary Metal-Oxide Semiconductor) camera while the HWP continuously rotated to determine the angle at which the maximum scattering intensity is detected, which is defined as the zero degree of rotation. Starting from this reference angle, the scattering intensity is recorded by the spectrometer, with the HWP then being rotated by $\pi/36$ rad (5°) between successive acquisitions, therefore rotating the input linear polarization of an angle of $\theta_{in} = 2\theta_{in,HWP} = \pi/18$. The extracted values are presented in **Figure 3.D**, where blue and red represent the nanopillar monomer and dimer, respectively. The normalized scattered intensity $(I(\theta_{in})/I_{max,IN}) = I(\theta_{in})/I(0)$) exhibits a strong dependence on the input polarization for both the monomer and dimer and follows a Malus' law-like cosine square function $I(\theta_{in})/I(0) = a \cdot \cos^2(\theta_{in}) + b$ (but in this case, the solution is represented by the relative scattered light intensity[42]), $a$ quantifies the intensity difference between the minimum and maximum, and $b$ is a non-polarization-dependent factor that quantifies the non-polarized component of light scattered by the structure. This behavior can be attributed to the intensity distribution of the evanescent field propagating along the surface of the TNF. The coupling efficiency between the TNF and the plasmonic structure is determined by the intensity of the evanescent field at the fiber surface where it interacts with the structure. Due to the asymmetric distribution of the electromagnetic field, the coupling efficiency varies accordingly, reaching its maximum when light in input is polarized along the *y*-axis and its minimum along the *x*-axis. This is confirmed numerically (continuous line), where both the structures show similar behavior for the normalized scattered intensity, with $b_{monomer,IN} = 0.045$ and $b_{dimer,IN} = 0.05$. The extracted experimental values are in good agreement with the numerical results for the monomer, which shows $b_{monomer,IN} = 0.095$, while the dimer shows a higher $b_{dimer,IN} = 0.21$, even if with a clear dependence from the input polarization.

With reference to the configuration depicted in **Figure 3.E**, to optimize the coupling with the nanostructures, a 785 nm beam linearly polarized along the *y*-axis was coupled into the TNF at the entrance upon passing through WP, set at the angle that previously gave $I_{max,IN}$. The scattered light then passed through a collection half-wave plate and a polarizer, transmitting light polarized along the *z*-axis, ensuring that the detector captured most of the scattered light. The measured polarization of the scattered light from the BEBID nanostructures and the numerical data are presented in **Figure 3.F**. While both structures show evident polarization of the scattered signal, the emergence of the hybridized modes in the dimer gap results in a stronger polarization, showing a numerical $b_{dimer,OUT} = 0.03$, and an experimental $b_{dimer,OUT} = 0.055$, while the monomer shows a numerical $b_{monomer,OUT} = 0.15$ and an experimental $b_{monomer,OUT} = 0.22$. The measurements indicate that the scattered signal is predominantly linearly polarized along the *z*-axis, corresponding to the TNF's longitudinal axis. At the same time, the component along the *x*-axis contributes minimally to the overall scattered intensity, while the contribution along the *y*-axis is not accessible during the measurement. Given that linearly polarized light along the *y*-axis was coupled into the TNF, the *z*- and *y*-components of the electric field were primarily coupled to the plasmonic nanostructure. However, the *y*-component is not observable from the top of the nanopillar, resulting in the scattered signal in the *xz*-plane being polarized along the *z*-axis. The linear polarization state of the scattered signal can be quantified using the Degree of Linear Polarization (DoLP), defined as:

$$DoLP = \frac{I_{max,OUT} - I_{min,OUT}}{I_{max,OUT} + I_{min,OUT}}, \qquad (2)$$

where $I_{max,OUT}$ and $I_{min,OUT}$ represent, respectively, the maximum electric field intensity, obtained for a polarization aligned along the *z*-axis on the *zx*-plane, and the minimum intensity, obtained for a linear polarization perpendicular to the *z*-axis. From the extracted data, *DoLP* can be calculated: the monomer on the TNF exhibits a *DoLP* of 0.65, whereas the dimer on the TNF shows a higher *DoLP* of 0.92, both polarized along the *z*-axis.

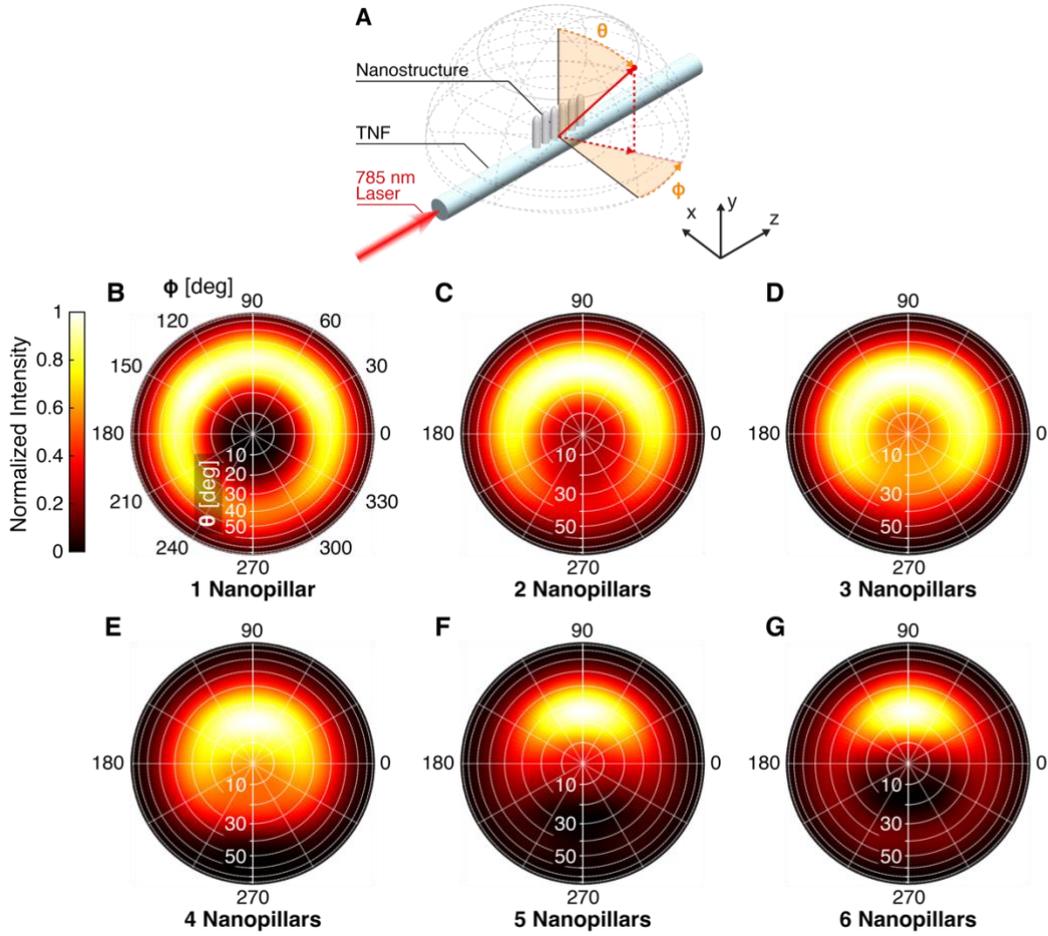

*Figure 4. Normalized scattered far field intensity ($|E|^2$) on a 2π hemisphere enclosing the nanopillars. A) Definition of the coordinates system. B-G) Polar maps of the far-field intensity distribution as the number of nano-pillars increases from 1 nanopillar to 6 nanopillars.*

Finally, to investigate the potential for ordered nanopillar arrays to control the direction of the scattered electromagnetic fields, we calculated the scattered far-field intensity ($|E|^2$) on a 2π hemisphere enclosing progressively increasing numbers of nanopillars, from one to six. Each structure was excited by the fundamental mode of the TNF (λ = 785 nm) propagating through the fiber. **Figure 4** displays the scattered far-field intensity on standard spherical coordinates, with each configuration normalized to its maximum intensity. In **Figure 4.A**, the coordinate system is defined: the polar angle θ extends from the axis normal to the TNF surface, aligned with the nanopillars' axis (the inner circles correspond to θ values from 0° to 90°). The azimuthal angle ϕ rotates counterclockwise on the surface plane (ϕ = 90° is taken as the forward-propagating direction).

With a single nanopillar (**Figure 4.B**), scattering occurs across all azimuthal directions, and regions with over 70% of the maximum intensity span approximately 300° around the nanopillar, with a central orientation in the forward-propagating direction. When two nanopillars are arranged in a nanoantenna configuration (**Figure 4.C**), the high-intensity scattering range above 70% of the maximum narrows to less than 240°. It shows a substantial reduction in the back-propagating component (between 240° and 300° azimuth). This effect becomes more pronounced as the number of pillars increases, as shown in **Figures 4.D-G**, where the azimuthal distribution of the scattered field narrows into a forward-directed lobe. By the six-nanopillar configuration, this range reduces to approximately 90° in the forward-propagating direction. Numerical simulations of an increased number of pillars (**Supplementary Figure 14.A-C** for 7, 9, and 10 pillars) show that increasing the number of pillars in the array does not significantly reduce the azimuthal angle range, while the arising of a second scattering lobe is observed. The second lobe is normal to the fiber, so it is expected to reduce the overall portion of light redirected toward the fiber collection direction. These results demonstrate that the far-field scattering directivity is strongly influenced by the number of nanopillars in the array, with increased forward-directional control as the array size grows. Furthermore, they indicate that the far-field emission direction can be tuned by adjusting the geometric arrangement of the nanopillar array[43].

## 4 Discussion and conclusions

In this work, we introduced and validated Blurred Electron Beam Induced Deposition (BEBID) as a novel bottom-up approach for the deterministic fabrication of plasmonic nanostructures directly onto suspended, single-mode tapered optical nanofibers (TNFs). Building on conventional Electron Beam Induced Deposition (EBID), this method relies on the deliberate and controlled defocusing of the electron beam to minimize the mechanical stress exerted on the nanometric dielectric fiber. This approach enables the deposition of platinum nanopillars with sub-20 nm spatial precision and high structural fidelity in a single-step process. Unlike conventional Focused EBID (FEBID), where feature resolution is defined by the beam spot size, BEBID allows variation in feature size by controlling the extent of the beam blurring. However, the minimum achievable feature size remains dependent on the characteristics of the substrate. For instance, on the TNFs studied in this work, beam blurring below 50 nm proved insufficient to alleviate beam-induced stress on the fiber. Future advances in TNF handling and packaging are expected to overcome this limitation, broadening the applicability of BEBID to controlled nanofabrication on fragile substrates and opening new pathways for robust, high-precision plasmonic architectures beyond the limitations of conventional technologies.

Through comprehensive optimization of fabrication parameters – including beam current, dwell time, and post-deposition plasma treatment – we achieved precise control over both geometry and composition. The resulting structures exhibited excellent intra-sample diameter uniformity (below 1%) and high inter-sample variation (around 8%). Energy-dispersive X-ray spectroscopy (EDX) and SEM inspection confirmed that parameter optimization and post-processing plasma oxygen purification yielded platinum-rich structures (up to ~65% Pt by volume) with reduced carbon content, enhancing their plasmonic properties. Optical characterization of single nanopillars (monomers) and coupled dimer structures on the TNF revealed polarization-dependent scattering responses and a pronounced redshift in the extinction spectra of dimers, attributed to hybridized plasmonic modes in the nanogap. Experimental measurements, including polarization-resolved scattering maps and broadband extinction spectra, closely matched FDTD simulations, confirming both the resonance tuning and the polarization properties of the scattered field. Simulations further predicted that increasing the number of nanopillars in ordered arrays narrows the far-field angular scattering distribution, progressively enhancing forward directivity along the fiber axis.

Following this proof-of-concept demonstration, several important directions emerge, warranting future studies to explore open questions raised by this work. While the influence of beam parameters on nanostructure composition is qualitatively clear, a systematic study of the composition variation mapping the interplay between accelerating voltage, beam current, and dwell time is needed to establish robust fabrication protocols. Additionally, although the versatility to switch deposited material, which can be modified by selecting a different precursor, is certainly an advantage, appropriate "purification" techniques, equivalent to plasma oxygen exposure for the precursor used in this work, will need to be investigated for each case. Purification of structures deposited via EBID is a topic widely addressed by the community[46], and elevated purification levels have been observed for multiple metals, using (and not limited to) oxygen[47] or water vapor[48] exposure simultaneously to the EBID process, or post-exposure to an electron beam[49] or a laser beam[50]. In the context of coupling BEBID-fabricated nanostructures with solid-state single-photon emitters to improve single-photon emission properties, precise relative positioning between the emitters and the plasmonic structures becomes a central challenge. In this regard, while, on the one hand, advanced emitter placing techniques such as atomic force microscopy pick-and-place[51,52] could achieve precise positioning of the emitter in close proximity to a deposited nanostructure, on the other hand, it would be desirable to place the emitter first and subsequently grow the nanostructure around it via BEBID, to exploit the highly deterministic positioning of the technique, so as to achieve the best relative positioning of the two components. However, this approach requires careful investigation of the effect of EBID on the emitter's optical properties.

Looking ahead, building on the main feature of the technology shown here, the precise control of the placement, spacing, and geometry of the various nanopillars, together with the possibility of varying the relative composition of each pillar via fabrication parameters, suggests the possibility of exploring the optical behavior of non-periodic arrays, including Yagi-Uda nanoantennae[44,45], which will increase the redirecting effect of the electromagnetic field, emitter-coupled configurations or even hybrid arrays, in which the optical properties of each element of the array will be defined dynamically. Altogether, BEBID emerges as a powerful and versatile platform for engineering light–matter interactions at the nanoscale, with significant potential implications for quantum photonics, directional emission control, and ultra-sensitive plasmonic sensing.

# 5 Materials and methods

## 5.1 TNF preparation

The TNFs were manufactured from a commercial optical fiber (Thorlabs 780HP) by the "heat-and-pull" method described in Ref.[18], with a custom setup reported in **Supplementary Figure S5**, designed to achieve high-transmission nanoscale fiber diameters. This setup includes an oxyhydrogen flame capable of heating fused silica to its softening point (1585 °C) and two translation stages, each equipped with a clamp to secure and pull the fiber ends. The flame is generated by a stoichiometric hydrogen-oxygen mixture, with hydrogen produced via water electrolysis. Precise flame control is achieved through two mass flow controllers, providing stable and reproducible heating conditions. During the pulling process, a motorized stage (Thorlabs PT1-Z9) precisely positions the flame below the fiber. Initially, the flame heats the fiber to its softening point. The stage lifts the flame just before pulling ends to prevent excessive heating, ensuring a smooth tapering process.

The tapering process was designed to satisfy the adiabatic criterion[53], which prevents light from coupling from the fundamental mode $HE_{11}$ to higher-order modes. Using a MATLAB script, trajectories for the translation stages are calculated based on the desired taper profile, adiabatic criterion (both reported in **Supplementary Figure S6**), and target nanofiber diameter.

Throughout this process, a 785 nm laser (Thorlabs S1FC780) is transmitted through the fiber, with transmission monitored by a photodiode. This output is normalized to the initial transmission to ensure consistent light throughput. To verify the diameter reduction in real-time, a camera with a long-working distance objective provides high-magnification monitoring. The transmission curve recorded during the pulling process (**Supplementary Figure S7**) shows that the fiber maintained 99.7% of the signal after fabrication. Statistics of the fabricated fiber diameter and transmission are provided in **Supplementary Table S1**. Maintaining a dust-free environment is essential to prevent particulate adhesion, which can reduce transmission.

To enable stable nanostructure fabrication using an SEM, a custom fiber holder was designed for mounting the TNFs in the machine (**Supplementary Figure S8**). However, due to manual assembly, strong vibrations were observed during SEM scanning. To mitigate this issue, the TNF was stretched an additional length $\Delta L$ after the pulling process, and before mounting it on the holder. Empirical observations showed that by increasing $\Delta L$ between 75 µm and 125 µm, the vibrations during electron beam scanning became negligible, while further increasing $\Delta L$ would increase the risk of breaking the fiber. This tensioning allowed for stable SEM imaging of the fiber at the magnification required for precise fabrication, as demonstrated in **Supplementary Figure S9**. After tensioning, the fibers were glued to the holder and stored in a clean environment.

## 5.2 Blurred EBID

Nanostructure fabrication was performed using a dual-beam Focused Ion Beam/Scanning Electron Microscope (FIB/SEM), FEI Helios NanoLab 600i, equipped with a gas injection system (GIS). Electrons were generated through a Schottky field emission gun. For the different fabrication used in the paper, the accelerating voltage was set at 1 kV, 3 kV, 5 kV, and 10 kV, while beam currents of 5.4 pA, 23 pA, 86 pA, and 340 pA were used. Higher current values negatively impacted the TNF's structural integrity. For the final fabrication of nanopillars on the TNF, a setting of 1 kV and 86 pA was used for both imaging and deposition.

Achieving precise focus on the TNF is essential for fabrication accuracy. To achieve optimal focus before fabrication, we recommend adjusting imaging conditions on an alignment sample (or on the fiber holder if no alignment sample is available) to minimize exposure of the nanometric TNF segment. Prolonged exposure, especially at high voltages or beam currents, can cause damage, including fiber rupture. After focusing, the nanometric section of the TNF was positioned under the electron beam at high magnification, ensuring the fiber's diameter occupied approximately one-fifth of the field of view's width.

Fabrication commenced once the chamber reached a target vacuum level of $\sim 5 \times 10^{-7}$ mbar. The GIS needle was then introduced in the chamber, above the TNF, allowing the Pt precursor MeCpPtMe$_3$ (*Trimethyl [(1,2,3,4,5-ETA.)-1 Methyl 2, 4-Cyclopentadien-1-YL] Platinum*) to flow. After initiating the gas flow, we allowed the chamber pressure to stabilize at $0.8$–$1.2 \times 10^{-5}$ mbar, maintaining the precursor temperature at $\sim$49 °C to reduce variations in the growth rate.

Upon completion, we paused to allow the vacuum to return to the level measured before opening the GIS, preventing unintended deposition from residual precursor gas exposure to the SEM beam.

## 5.3   EDX analysis

Energy Dispersive X-Ray Spectroscopy measurements were performed on a test sample using a Thermo Scientific Phenom Pro G6 Desktop SEM equipped with an EDX detector. The test sample consisted of a 170 µm-thick borosilicate glass cover (Ted Pella), coated by a 100 nm 99.99% pure gold (Kurt J. Lesker) layer deposited by electron beam evaporation (Thermoionics Laboratory Inc.). The gold layer is necessary to perform subsequent SEM imaging and fabrication. The test sample was processed with a FEI Helios Nanolab 600i Scanning Electron Microscope / Focused Ion Beam dual beam system. Firstly, a $5 \times 4$ matrix of $3 \times 3$ µm$^2$ squares was milled with FIB milling (30 kV, 0.41 nA) for a depth of $\sim$1 µm to remove the gold and reach the underneath glass. Subsequently, EBID was used to realize a series of 5 cubes of $500 \times 500 \times 500$ nm$^3$, each at the center of a previously milled area, using different electron beam current and an accelerating voltage of 1 kV. EDX spectra were acquired using a 15 kV electron beam in fixed point modality and an exposure time of 30 s, and were analyzed with the Thermo Scientific Phenom ProSuite Software to extract the atomic percentage of Pt relative to C.

## 5.4   Optical characterization

For the optical characterization, the nanostructure-decorated TNFs' ends were cleaved and connected to two temporary FC/PC connectors (Thorlabs BFT1 terminator and B30126C3 connector) and mounted in a custom-made sealed Plexiglas box to prevent dust deposition during the measurements. Light scattered from the structures was collected through a broadband optical window (Thorlabs WG41010R). The optical characterization setup comprised two light injection branches (one for broadband light injection and one for polarization-resolved monochromatic light injection) and a common detection path. For the broadband light injection branch, a supercontinuum laser (NKT Photonics SuperK Compact, fiber coupled) was intensity-filtered by a variable neutral density filter (Thorlabs NDC-100C-4M), collimated, and focused on the entrance of a short single-mode patch cord, made from the same fiber used for TNFs, through an FC/PC triplet fiber collimator (Thorlabs TC12FC-780). For the polarization-resolved monochromatic light injection branch, a $\lambda = 785$ nm (Toptica Photonics DL PRO 780 S) laser passed through a half-wave plate (Thorlabs WPHSM05-780 mounted in Thorlabs RSP1X15 indexing rotation mount) and a variable neutral density filter before being coupled to the distal end of the same short patch cord through another similar collimator. The short patch cord was then connected to one end of the TNF. In the common detection path, light scattered by the structures and passed through the optical window is collected by an aspheric lens (Thorlabs A220TM-B) mounted on a 3-axis piezoelectric stage (Piezosystem Jena TRITOR 102 T-405-00). In the broadband measurements, the collected signal was sent toward a moveable 50:50 beam splitter that temporarily sent a fraction of it to a sCMOS camera (Hamamatsu Orca Flash 4) to image the structure. By removing the beam splitter, the signal was entirely sent toward a plano-convex lens (Thorlabs LA1131-B), which then focused it at the end facet of a multimode 0.22 NA, 200 µm core diameter patch cord (Thorlabs M25L05), used to route the signal towards a spectrometer (Princeton Instruments Acton Series SP2500i, equipped with a Pixis 400 Camera) for spectral recording. In the polarization dependence measurements, the detection path is slightly modified by adding a half-wave plate and a polarizer (Thorlabs LPVIS100-MP2) with its T-axis aligned with the *z*-direction, after the objective lens.

## 5.5 Numerical simulations

A Finite Difference Time Domain model (FDTD, Ansys Lumerical Solutions) was developed to simulate the nanostructure and assess both scattering spectra and electromagnetic field polarization under the influence of the TNF's fundamental mode propagation. To accurately represent the fabricated structure, the TNF is modeled as a silica ($SiO_2$) cylinder with a refractive index $n_{TNF} = 1.46$, while the nanopillars are represented as rounded-edge cylinders to approximate their physical shape. The carbon-platinum composite nanopillars' dielectric function and optical properties are modeled using Bruggemann's effective medium approximation over the relevant optical wavelength range. The surrounding medium is defined as free space with a refractive index of $n = 1$.

The simulation domain is defined as a large rectangular prism with perfectly matched layer (PML) boundary conditions in all directions, carefully sized to prevent interaction between the evanescent fields and the PML boundaries. A mesh override is applied to the TNF and nanopillars for higher precision, while the rest of the domain employs Lumerical's built-in nonuniform mesh algorithm with a mesh accuracy setting of 4. To compute the mode profile for the TNF, we use the Ansys MODE eigensolver, and the resulting guided mode is injected into the TNF using a mode source in the FDTD model. This approach ensures accurate mode propagation and interaction with the nanopillars in the simulated environment.

# 6 Data Availability Statement

The data supporting the findings of this study are currently being curated and will be made available in a public repository upon publication of the article. Reasonable requests for access to the data prior to publication can be directed to the corresponding author.

# 7 Conflict of Interest Disclosure

The authors declare no conflict of interest.

# 8 Acknowledgments

A.Ba. acknowledges funding from the European Union's Horizon 2020 research and innovation program under the Marie Sklodowska-Curie grant agreement (#101106602). A.Ba., M.D.A., M.F.K., C.D., F.P., M.D.V., and A.Br. also acknowledge funding from the European Union's Horizon 2020 research and innovation program under a grant agreement (#828972). M.D.V., and F.P. acknowledge funding from the European Union's Horizon Europe under a grant agreement (#101125498). A.Br. is a member of the Institut Universitaire de France (IUF). A.Ba. and M.D.A. contributed equally to this work and are co-first authors. M.D.V., F.P., and A. Br contributed equally to this work and are co-last authors.

# Deterministic Bottom-Up Fabrication of Plasmonic Nanostructures on Optical Nanofibers via Blurred Electron Beam Deposition
## SUPPLEMENTARY MATERIAL

Antonio Balena[1,*], Marianna D'Amato[1], Muhammad Fayyaz Kashif[2,3], Chengjie Ding[1], Massimo De Vittorio[3,4], Ferruccio Pisanello[3], Alberto Bramati[1]

[1] *Laboratoire Kastler Brossel, Sorbonne Université, CNRS, ENS-PSL Research University, Collège de France, France*
[2] *Università degli Studi di Napoli Federico II – Department of Electrical Engineering and Information Technologies, Italy*
[3] *Istituto Italiano di Tecnologia - Center for Biomolecular Nanotechnologies, Italy*
[4] *Technical University of Denmark - Department of Health Technology Drug Delivery and Sensing, IDUN Section, Denmark*

[*] *Corresponding author:* antonio.balena@lkb.upmc.fr

## 10  Tapered Optical Nanofiber (TNF) modeling and realization

**Supplementary Figure S1** illustrates the effective refractive index $n_{eff}$ as a function of the TNF diameter (bottom x-axis) for the first four lower-order modes - $HE_{11}$, $HE_{21}$, $TM_{01}$, and $TE_{01}$ – considering $n_1 = 1.46$, $n_0 = 1.00$, and $\lambda = 785$ nm. The effective refractive index $n_{eff}$ is defined as $n_{eff} = \beta/k$, where $\beta$ represents the mode propagation constant and $k = 2\pi/\lambda$ is the wavenumber. The normalized frequency parameter $V$, is defined as:

$$V = \frac{2\pi}{\lambda}\sqrt{(n_1^2 - n_0^2)},$$

and it is displayed on the top x-axis, where *a* is the radius of the TNF, *λ is* the wavelength of interest, and $n_1$ and $n_0$ are the refractive indices of the cladding and surrounding air, respectively. In the nanofiber region, as the original core is melted away, the cladding now assumes the role of the core, while air acts as the surrounding medium.

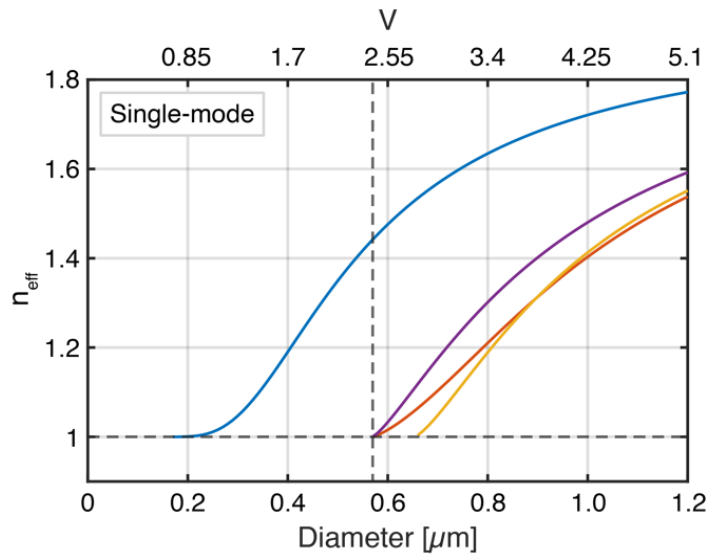

***Supplementary Figure S1. Effective refractive index $n_{eff}$ as a function of the TNF's diameter and the normalized frequency parameter V.** The boundary between single-mode and multi-mode regimes is marked by a vertical dashed line (V = 2.405). Below this threshold, only the fundamental mode $HE_{11}$ propagates inside the TNF.*

Each mode, except the fundamental $HE_{11}$ mode, can only propagate if $V$ exceeds a mode-specific cut-off value. For $V < 2.405$, only the $HE_{11}$ mode is supported, meaning the nanofiber operates in a single-mode regime.

## 10.1 Derivation of the electromagnetic field around the TNF

The fundamental mode $HE_{11}$ of an TNF is given by the three components of the electric and the magnetic fields $E_x$, $E_y$, $E_z$ and $H_x$, $H_y$, $H_z$ in cylindrical coordinates $(r, \theta, z)$. At the air-cladding interface, given the radius $a$ of the TNF, these components can be expressed as follows.

For $r < a$ (inside the fiber):

$$E_x = -jA\beta \frac{a}{u}\left[\frac{(1-s)}{2}J_0\left(\frac{u}{a}r\right)\cos\psi - \frac{(1+s)}{2}J_2\left(\frac{u}{a}r\right)\cos(2\theta+\psi)\right],$$

$$E_y = jA\beta \frac{a}{u}\left[\frac{(1-s)}{2}J_0\left(\frac{u}{a}r\right)\sin\psi + \frac{(1+s)}{2}J_2\left(\frac{u}{a}r\right)\sin(2\theta+\psi)\right],$$

$$E_y = AJ_1\left(\frac{u}{a}r\right)\cos(\theta+\psi),$$

$$H_x = -jA\omega\varepsilon_0 n_1^2 \frac{a}{u}\left[\frac{(1-s_1)}{2}J_0\left(\frac{u}{a}r\right)\sin\psi + \frac{(1+s_1)}{2}J_2\left(\frac{u}{a}r\right)\sin(2\theta+\psi)\right],$$

$$H_y = -jA\omega\varepsilon_0 n_1^2 \frac{a}{u}\left[\frac{(1-s_1)}{2}J_0\left(\frac{u}{a}r\right)\cos\psi - \frac{(1+s_1)}{2}J_2\left(\frac{u}{a}r\right)\cos(2\theta+\psi)\right],$$

$$H_y = -A\frac{\beta}{\omega\mu_0}sJ_1\left(\frac{u}{a}r\right)\sin(\theta+\psi).$$

For $r > a$ (outside the fiber):

$$E_x = -jA\beta \frac{aJ_1(u)}{\omega K_1(\omega)}\left[\frac{(1-s)}{2}K_0\left(\frac{\omega}{a}r\right)\cos\psi + \frac{(1+s)}{2}K_2\left(\frac{\omega}{a}r\right)\cos(2\theta+\psi)\right],$$

$$E_y = jA\beta \frac{aJ_1(u)}{\omega K_1(\omega)}\left[\frac{(1-s)}{2}K_0\left(\frac{\omega}{a}r\right)\sin\psi - \frac{(1+s)}{2}K_2\left(\frac{\omega}{a}r\right)\sin(2\theta+\psi)\right],$$

$$E_z = A\frac{J_1(u)}{K_1(\omega)}K_1\left(\frac{\omega}{a}r\right)\cos(\vartheta+\psi),$$

$$H_x = -jA\omega\varepsilon_0 n_0^2 \frac{aJ_1(u)}{w K_1(\omega)}\left[\frac{(1-s_0)}{2}K_0\left(\frac{\omega}{a}r\right)\sin\psi - \frac{(1+s_0)}{2}K_2\left(\frac{\omega}{a}r\right)\sin(2\theta+\psi)\right],$$

$$H_y = -jA\omega\varepsilon_0 n_0^2 \frac{aJ_1(u)}{w K_1(\omega)}\left[\frac{(1-s_0)}{2}K_0\left(\frac{\omega}{a}r\right)\cos\psi + \frac{(1+s_0)}{2}K_2\left(\frac{\omega}{a}r\right)\cos(2\theta+\psi)\right],$$

$$H_z = -A\frac{\beta}{\omega\mu_0}s\frac{J_1(u)}{K_1(\omega)}K_1\left(\frac{\omega}{a}r\right)\sin(\theta+\psi),$$

Where $\psi$ is the polarization angle, $s = [(\omega)^{-2} + (u)^{-2}]/[J_1'(u)/uJ_1(u) + K_1'(\omega)/\omega K_1(\omega)]$, $s_1 = \beta^2 s/(k^2 n_1^2)$, $s_0 = \beta^2 s/(k^2 n_0^2)$, $\beta$ is the propagation constant, $u = a\sqrt{n_1^2 k^2 - \beta^2}$ and $w = a\sqrt{\beta^2 - n_0^2 k^2}$ ($u^2 + w^2 = V^2$, where $V$ is the normalized frequency). $J_n$ and $K_n$ are the Bessel functions of the first type and the modified Bessel functions of the second type, respectively. The single quote mark stands for derivative. **Supplementary Figures S2.A-F** show, respectively, the components $E_x$, $E_y$, $E_z$, $H_x$, $H_y$, and $H_z$ of the electric and magnetic fields of the fundamental mode $HE_{11}$ calculated for $\lambda = 785$ nm guided light linearly polarized along the y-axis for a TNF with a diameter of 360 nm, while **Supplementary Figure S3.A** illustrates the normalized electric field intensity profile ($\lambda$ = 785 nm) for the fundamental mode $HE_{11}$ as a function of the TNF waist diameter $d$, with light linearly polarized along the horizontal axis ($x = 0$). The center of the TNF is at $y = 0$, and $y = \pm d/2$ are the upper and lower TNF boundaries. As $d$ decreases from right to left, the evanescent field outside the fiber – observed for $y > d/2$ or $y < -d/2$ – gradually increases, becoming the major contribution, particularly for sub-wavelength $d$. However, when the fiber diameter becomes too small, light confinement to the fiber surface is significantly reduced. The strongest evanescent field at the fiber surface occurs at a fiber waist diameter of $d = 360$ nm, with the corresponding calculated electric field intensity $|E|^2$ on the transversal plane shown in **Supplementary Figure S3.B** along with detailed equations. A representative SEM Micrograph of a TNF's minimum waist section is shown in Supplementary **Figure S4.A**, along with an extracted grayscale profile from which it can be observed that the TNF features a waist of $d = 360$ nm, that is maintained along a length of 3 mm. For diameters in the range $d \epsilon$ [320, 400] nm, the evanescent field intensity at the fiber surface remains above 90% of the maximum (**Supplementary Figure S4.B**), thus defining a tolerance that accounts for slight diameter variations during fabrication.

.

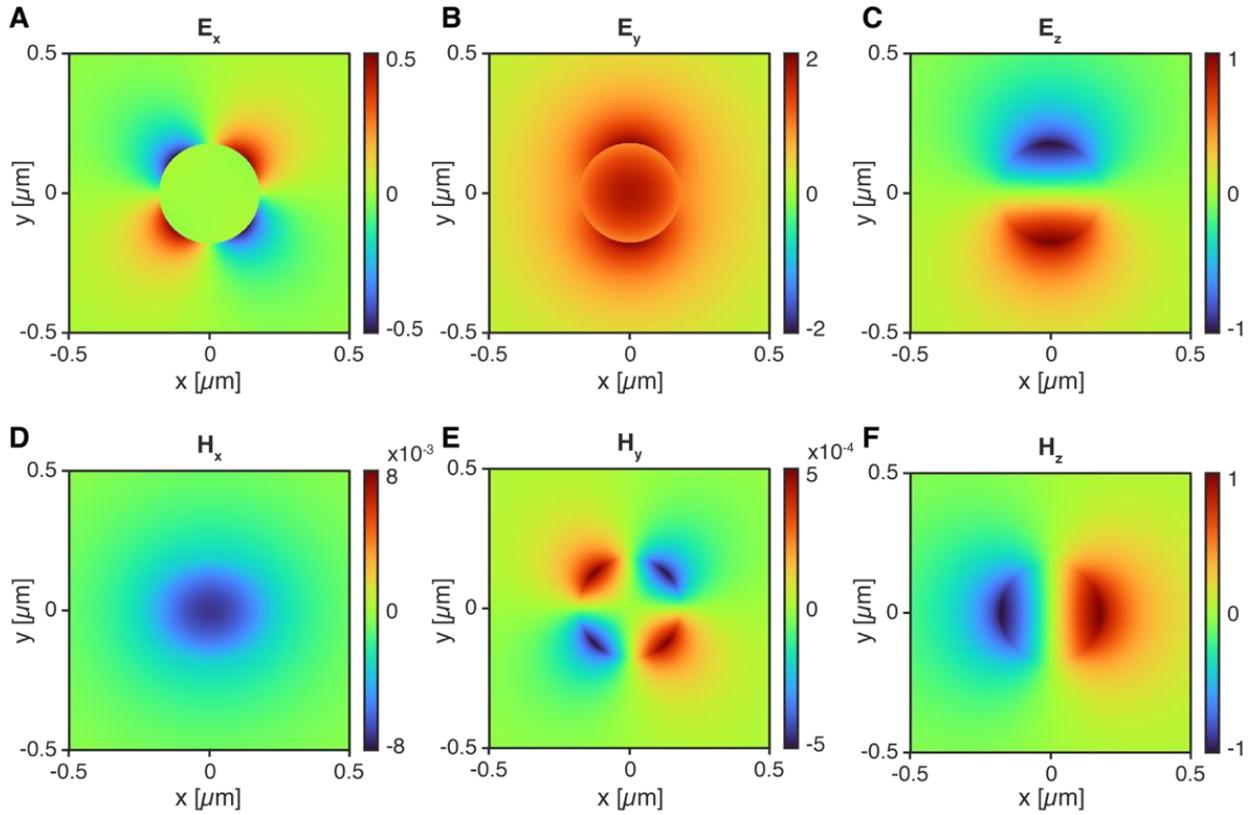

***Supplementary Figure S2. Fundamental mode $HE_{11}$ structure. A-F)*** *The components $E_x$, $E_y$, $E_z$, $H_x$, $H_y$, and $H_z$, respectively, of the electric and magnetic fields of the fundamental mode $HE_{11}$. The components are simulated for a 360 nm-diameter TNF and in the case of a 785 nm guided light with linear polarization along the y-axis using the equations in section S1.1.*

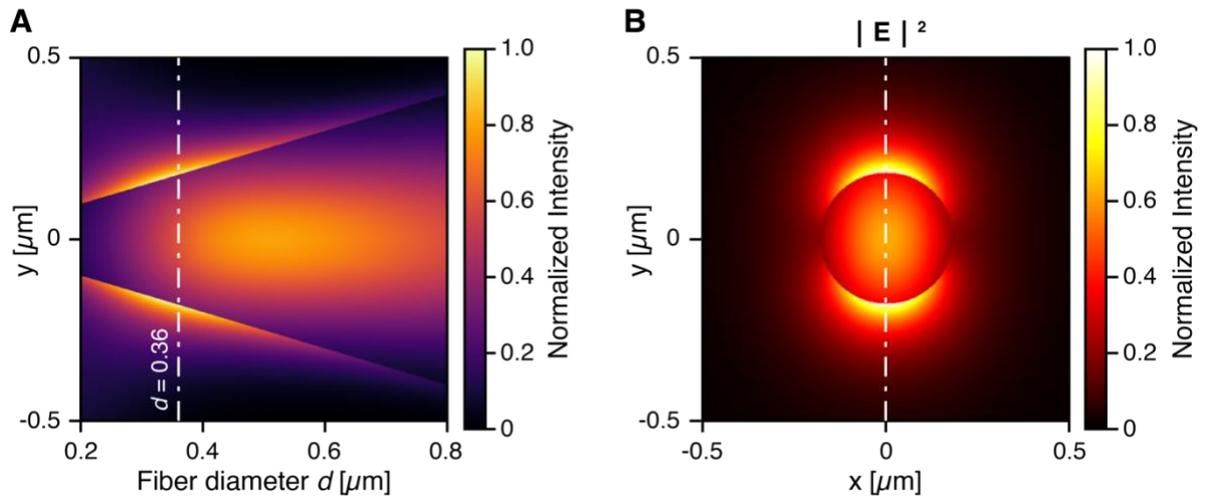

*Supplementary Figure S3. electric field intensity calculations.* **A**) Map of the electric field intensity profile along the central axis of the TNF transversal plane versus the TNF diameter for λ = 785 nm. The dot-dashed white line highlights the maximum intensity profile, obtained for a waist diameter of d = 360 nm. **B**) The normalized electric field intensity distribution on the transversal plane for d = 360 nm and λ = 785 nm.

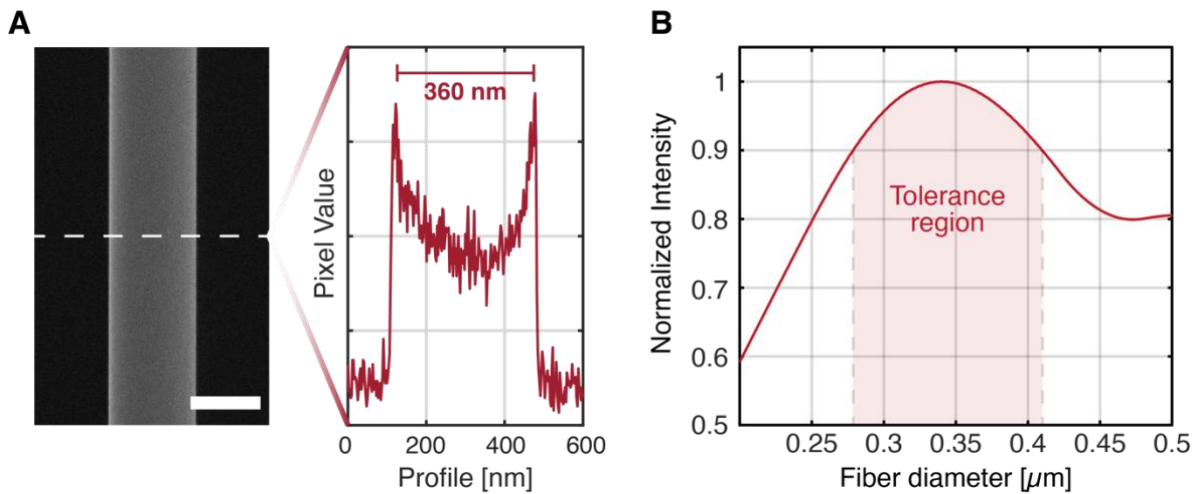

*Supplementary Figure S4. Definition of the diameter tolerance region.* **A**) Scanning Electron Micrograph of the minimum waist section of a tapered nanofiber and extracted grayscale values profile along the white dashed line. The scale bar is 300 nm. **B**) Plot of the maximum electromagnetic field intensity for r > a (evanescent field outside the fiber) versus the nanofiber diameter 2a, normalized at the highest value. The shaded area identifies the range of diameters in which the maximum intensity remains above 90% of the absolute maximum, thereby defining a fabrication tolerance region for the nanofiber.

## 10.2 TNF's fabrication and optimization

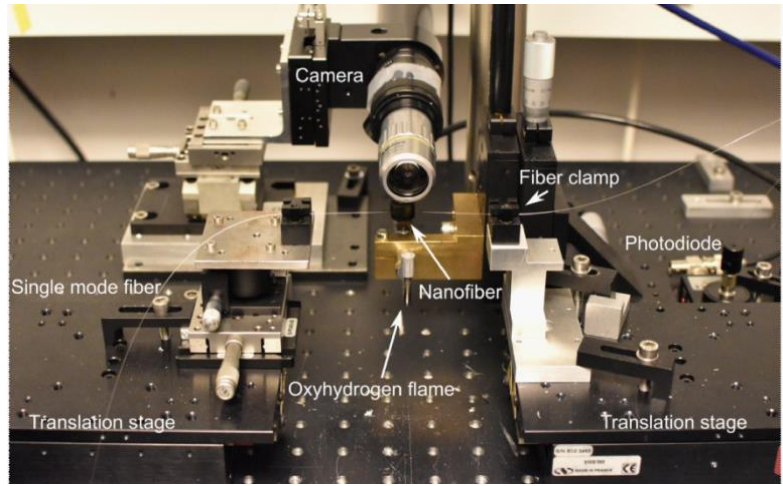

***Supplementary Figure S5. Custom heat-and-pull setup.*** *The two translation stages follow a precisely calculated pulling trajectory while the oxyhydrogen flame heats the single-mode fiber segment between the two fiber clamps to obtain a nanofiber. A camera continuously monitors the process, and a photodiode retrieves the fiber transmission in real time.*

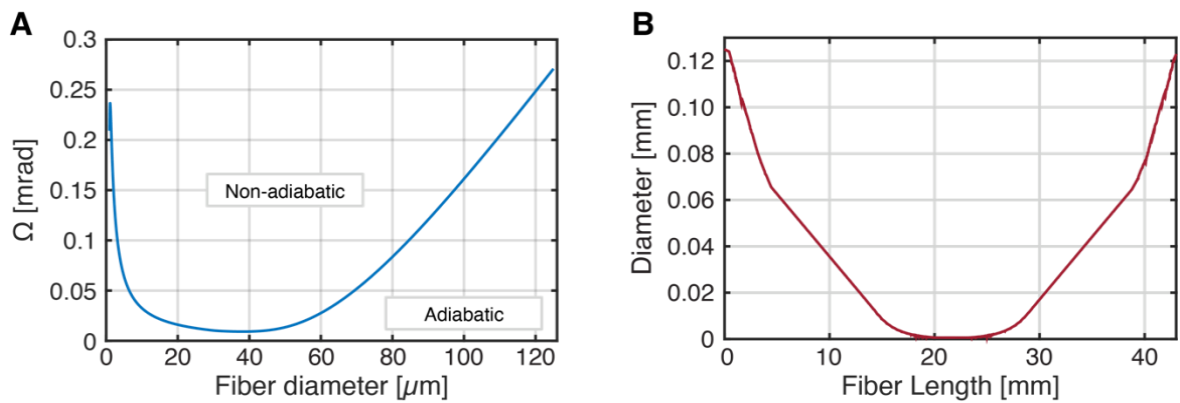

***Supplementary Figure S6. Adiabatic condition.*** *A) Adiabatic criterion calculated for a Thorlabs HP780 optical fiber, a diameter $d = 360$ nm, a nanofiber length of $l = 3$ mm, and a working wavelength of $\lambda = 785$ nm. B) Tapered nanofiber adiabatic profile calculated to obtain adiabatic transmission throughout the entire fiber, satisfying the conditions defined in Panel A.*

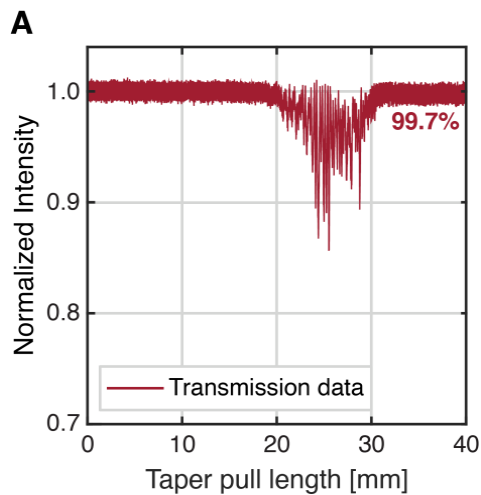

***Supplementary Figure S7.*** *Exemplary plot of the measured transmission during the heat-and-pull process, for which the advancement is quantified by the taper pull length. Unitary transmission represents the transmission of the fiber before starting the heat-and-pull.*

| Sample | Transmission [%] | Waist diameter $d$ [nm] |
|---|---|---|
| Fiber 1 | 99.5 | 380 |
| Fiber 2 | 98.7 | 357 |
| Fiber 3 | 99.5 | 323 |
| Fiber 4 | 99.6 | 364 |
| Fiber 5 | 97.5 | 359 |

*Supplementary Table S1. Quality inspection of the fabricated TNFs. Transmission values are relative to the transmission intensity measured before the heat-and-pull process. Waist diameter is measured by Scanning Electron Microscope inspection.*

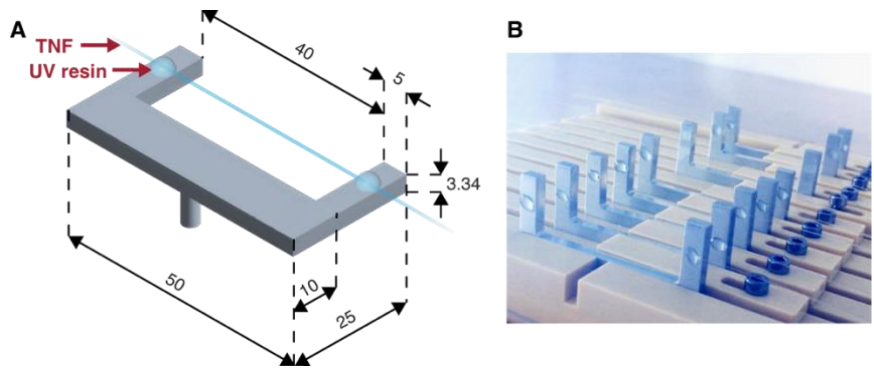

*Supplementary Figure S8. Custom holder. A) Sketch of the custom holder designed to insert the TNFs in the SEM machine. TNF not in scale. B) Photograph of prepared nanofibers glued to the custom holder and stored in a clean environment.*

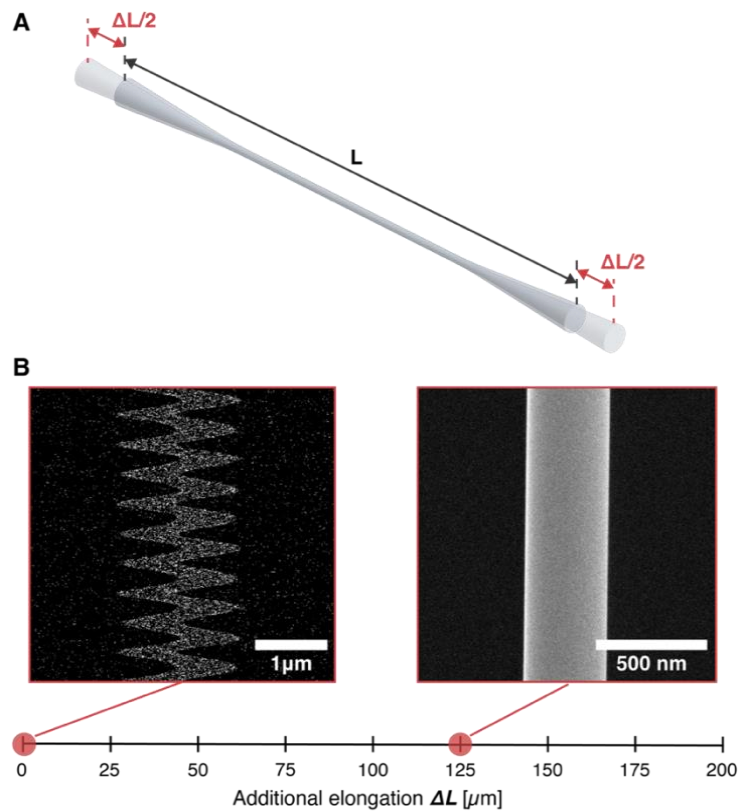

*Supplementary Figure S9. Additional elongation. A) Sketch of a TNF of length L at which an elongation of ΔL/2 on each end of the TNF is applied B) SEM micrographs of two d = 360 nm TNFs. On the left, no additional elongation has been applied after the pulling. On the right, a total additional elongation of L = 125 μm has been applied, thus enabling stable imaging of the TNF.*

# 11 Plasma Oxygen treatment

## 11.1 Cleaning of the nanofiber

Accumulation of dust on the nanofiber poses a significant problem, as it leads to a drastic reduction in transmission, often dropping below 10% [*M. Fujiwara, K. Toubaru, and S. Takeuchi, "Optical transmittance degradation in tapered fibers," Opt. Express 19, 8596-8601 (2011)*]. To address this issue, we employed a Plasma Oxygen treatment. This method, commonly used to clean macroscopic flat substrates, is highly effective at removing contaminants at the nanoscale compared to traditional wet cleaning methods, such as solvent cleaning.

**Supplementary Figure S10** showcases the effectiveness of this treatment on a nanofiber: the upper panel shows the effect of light scattering from dust when a laser is directed into a nanofiber. Under these conditions, it becomes nearly impossible to distinguish between the nanostructure and the dust, as both act as scattering centers. The bottom panel displays the result after plasma cleaning, where the nanostructure becomes clearly visible and the nanofiber is completely free of dust.

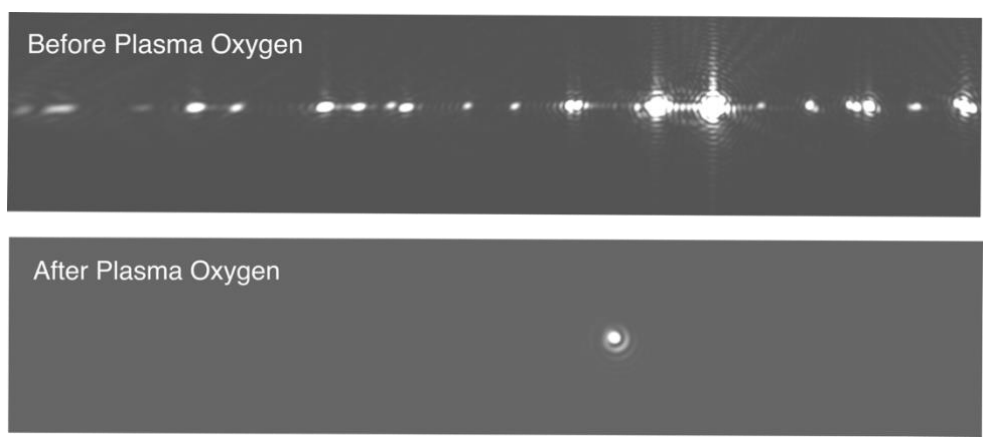

*Supplementary Figure S10. Plasma Oxygen cleaning effect. (upper panel) Before Plasma Oxygen treatment light scattering from dust heavily masks the scattering from the fabricated structures when a laser is sent into the nanofiber. (bottom panel) After Plasma Oxygen treatment the dust is removed from the nanofiber and only the scattering from the nanostructure is visible.*

## 11.2 Morphological effects of Plasma Oxygen

The Plasma Oxygen Treatment also leads to 'purification' effects on the composition of the EBID nanostructures, reducing the carbon concentration in the nanostructure, thus increasing the overall metal content. In turn, this treatment influences the dimensions of the nanostructures, potentially reducing both the diameter and the height of a few tens of nanometers. The effect has been characterized on a test planar sample (**Supplementary Figure S11**).

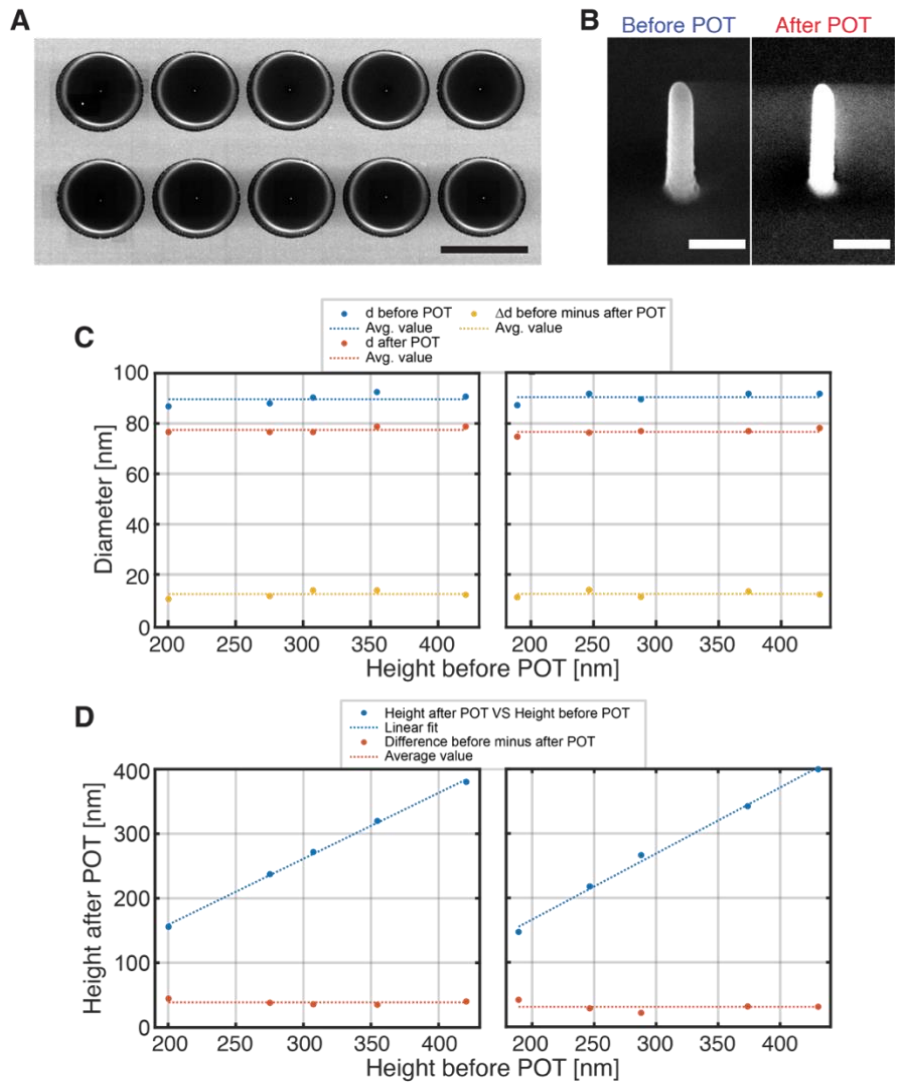

*Supplementary Figure S11. Plasma Oxygen morphological effect. A) SEM micrograph of the test structures consisting of 3 μm-diameter 1 μm-depth holes milled by FIB milling in a 100 nm-thick gold film deposited on a glass substrate. At the center of each hole, a single nanopillar is grown, increasing the nominal height from 200 to 400 nm, keeping a constant diameter of 100 nm. The scale bar is 3 μm.  B) Example SEM micrograph of a single pillar used for SEM inspection before and after the exposure to the plasma oxygen. Scale bars are 200 nm. C) Results of the SEM inspection to determine the diameter reduction after the plasma oxygen exposure on the upper row pillars (left) and the lower row ones (right). D) Results of the SEM inspection to determine the height reduction after the plasma oxygen exposure on the upper row pillars (left) and the lower row ones (right).*

## 12 Experimental and numerical supplementary material

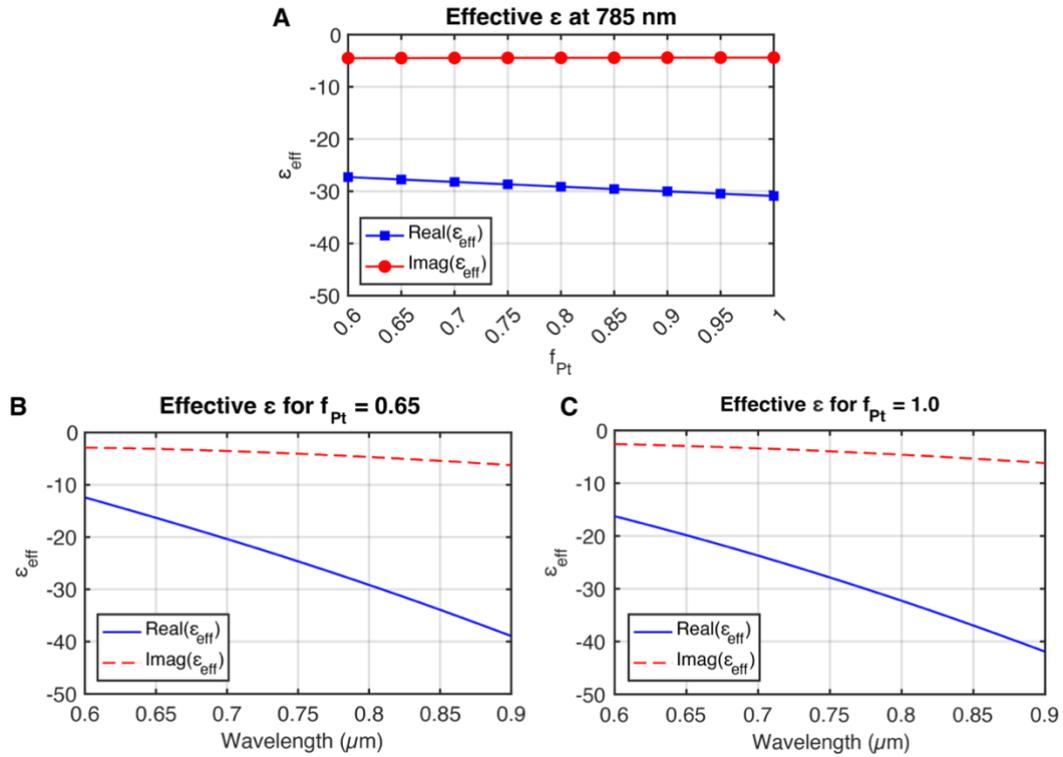

*Supplementary Figure S12. Effective medium theory calculations. A) Calculated effective dielectric permittivity at a fixed wavelength of λ = 785 nm for $f_{Pt} \in [0.6, 1.0]$ at steps of 0.05. B) Calculated effective dielectric permittivity for $f_{Pt} = 0.65$. C) Calculated effective dielectric permittivity for $f_{Pt} = 1.00$.*

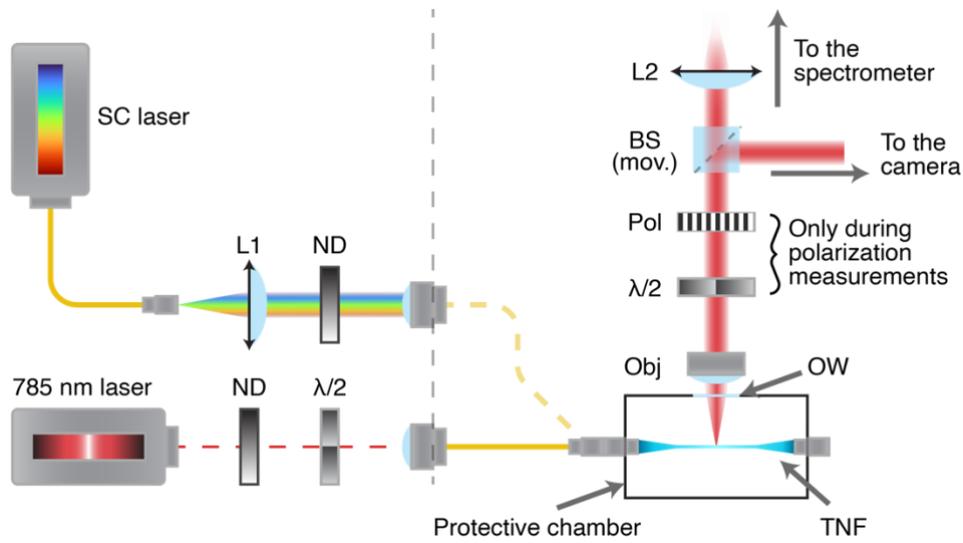

*Supplementary Figure S13. Sketch of the measurement setup. Two laser sources can be selectively sent in input inside the TNF: a supercontinuum source for the broadband scattering measurements, and a 785 nm laser for the polarization measurements. The TNF is kept inside an airtight Plexiglas container that prevents dust contamination during the experiments. The container is equipped with an optical window (Thorlabs WG41010R-B) with R < 0.5% in the wavelength range of interest. The signal is collected by an objective lens and sent toward the spectrometer. During the polarization measurement, a half waveplate and a polarizer are inserted in the optical path, as described in the main text. The sample is monitored thanks to an imaging branch enabled by a beam splitter that can be removed during the measurements, to maximize the collected signal.*

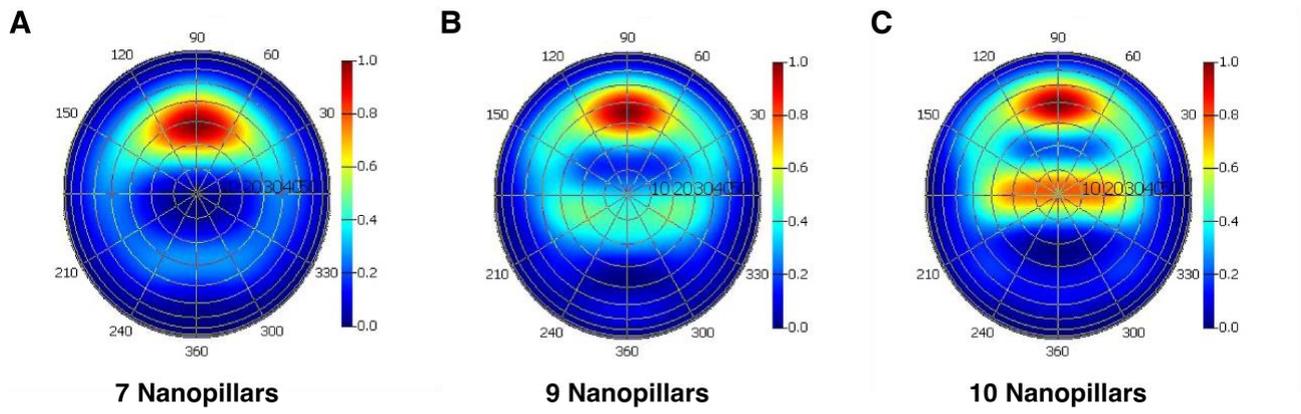

***Supplementary Figure 14. Normalized scattered far field intensity ($|E|^2$) on a $2\pi$ hemisphere enclosing the nanopillars. A-C)** Polar maps of the far-field intensity distribution of nano-pillar arrays containing, respectively, 7 nanopillars, 9 nanopillars, and 10 nanopillars.*